\documentstyle[12pt]{article}
\newcommand{\nc}{\newcommand}
\nc{\renc}{\renewcommand}

\nc{\etal}{\mbox{\it et al. }}
\nc{\ie}{{\it i.e.\ }}
\nc{\eg}{{\it e.g.}}

\renc{\thefootnote}{\arabic{footnote}}
\nc{\capt}[1]{{\bf Figure.} {\small\sl #1}}

\nc{\eqs}[2]{\mbox{Eqs.~(\ref{#1},\,\ref{#2})}}
\nc{\eq}[1]{\mbox{Eq.~(\ref{#1})}}

\nc{\figs}[2]{\mbox{Figs.~(\ref{#1},\,\ref{#2})}}
\nc{\fig}[1]{\mbox{Fig~.(\ref{#1})}}

\nc{\tag}[1]{\label{#1} \marginpar{{\footnotesize #1}}}
\nc{\mtag}[1]{\label{#1} \mbox{\marginpar{{\footnotesize #1}}}}
\renc{\baselinestretch}{1.2}
\newlength{\overeqskip}
\newlength{\undereqskip}
\nc{\be}[1]{\begin{equation} \mbox{$\label{#1}$}}
\nc{\bea}[1]{\begin{eqnarray} \mbox{$\label{#1}$}}
\nc{\Section}[2]{\section{#2}\label{#1}}
\nc{\Bibitem}[1]{\bibitem{#1}}
\nc{\Label}[1]{\label{#1}}

\nc{\eea}{\vspace{\undereqskip}\end{eqnarray}}
\nc{\ee}{\vspace{\undereqskip}\end{equation}}
\nc{\bdm}{\begin{displaymath}}
\nc{\edm}{\end{displaymath}}
\nc{\dpsty}{\displaystyle}
\nc{\bc}{\begin{center}}
\nc{\ec}{\end{center}}
\nc{\ba}{\begin{array}}
\nc{\ea}{\end{array}}
\nc{\bab}{\begin{abstract}}
\nc{\eab}{\end{abstract}}
\nc{\btab}{\begin{tabular}}
\nc{\etab}{\end{tabular}}
\nc{\bit}{\begin{itemize}}
\nc{\eit}{\end{itemize}}
\nc{\ben}{\begin{enumerate}}
\nc{\een}{\end{enumerate}}
\nc{\bfig}{\begin{figure}}
\nc{\efig}{\end{figure}}
\nc{\arreq}{&\!=\!&}
\nc{\arrmi}{&\!-\!&}
\nc{\arrpl}{&\!+\!&}
\nc{\arrap}{&\!\!\!\approx\!\!\!&}
\nc{\non}{\nonumber\\*}
\nc{\align}{\!\!\!\!\!\!\!\!&&}

\def\lsim{\; \raise0.3ex\hbox{$<$\kern-0.75em
      \raise-1.1ex\hbox{$\sim$}}\; }
\def\gsim{\; \raise0.3ex\hbox{$>$\kern-0.75em
      \raise-1.1ex\hbox{$\sim$}}\; }
\nc{\DOT}{\hspace{-0.08in}{\bf .}\hspace{0.1in}}
\nc{\Laada}{\hbox {$\sqcap$ \kern -1em $\sqcup$}}
\nc\loota{{\scriptstyle\sqcap\kern-0.55em\hbox{$\scriptstyle\sqcup$}}}
\nc\Loota{{\sqcap\kern-0.65em\hbox{$\sqcup$}}}
\nc\laada{\Loota}
\nc{\qed}{\hskip 3em \hbox{\BOX} \vskip 2ex}

\nc{\real}{{\rm I \! R}}
\nc{\Z}{{\sf Z \!\!\! Z}}
\nc{\complex}{{\rm C\!\!\! {\sf I}\,\,}}
\def\bigid{\leavevmode\hbox{\small1\kern-3.8pt\normalsize1}}
\def\id{\leavevmode\hbox{\small1\kern-3.3pt\normalsize1}}
\nc{\slask}{\!\!\!/}
\nc{\bis}{{\prime\prime}}
\nc{\pa}{\partial}
\nc{\na}{\nabla}
\nc{\ra}{\rangle}
\nc{\la}{\langle}
\nc{\goto}{\rightarrow}
\nc{\swap}{\leftrightarrow}

\nc{\EE}[1]{ \mbox{$\cdot10^{#1}$} }
\nc{\abs}[1]{\left|#1\right|}
\nc{\at}[2]{\left.#1\right|_{#2}}
\nc{\norm}[1]{\|#1\|}
\nc{\abscut}[2]{\Abs{#1}_{\scriptscriptstyle#2}}
\nc{\vek}[1]{{\rm\bf #1}}
\nc{\integral}[2]{\int\limits_{#1}^{#2}}
\nc{\inv}[1]{\frac{1}{#1}}
\nc{\dd}[2]{{{\partial #1}\over{\partial #2}}}
\nc{\ddd}[2]{{{{\partial}^2 #1}\over{\partial {#2}^2}}}
\nc{\dddd}[3]{{{{\partial}^2 #1}\over
	{\partial #2 \partial #3}}}
\nc{\dder}[2]{{{d #1}\over{d #2}}}
\nc{\ddder}[2]{{{d^2 #1}\over{d {#2}^2}}}
\nc{\dddder}[3]{{d^2 #1}\over
	{d #2 d #3}}
\nc{\dx}[1]{d\,^{#1}x}
\nc{\dy}[1]{d\,^{#1}y}
\nc{\dz}[1]{d\,^{#1}z}
\nc{\dl}[1]{\frac{d\,^{#1}l}{(2\pi)^{#1}}}
\nc{\dk}[1]{\frac{d\,^{#1}k}{(2\pi)^{#1}}}
\nc{\dq}[1]{\frac{d\,^{#1}q}{(2\pi)^{#1}}}

\nc{\cc}{\mbox{$c.c.$ }}
\nc{\hc}{\mbox{$h.c.$ }}
\nc{\cf}{cf.\ }
\nc{\erfc}{{\rm erfc}}
\nc{\Tr}{{\rm Tr\,}}
\nc{\tr}{{\rm tr\,}}
\nc{\pol}{{\rm pol}}
\nc{\sign}{{\rm sign}}
\nc{\bfT}{{\bf T }}
\def\GeV{{\rm\ GeV}}

\nc{\cA}{{\cal A}}
\nc{\cB}{{\cal B}}
\nc{\cD}{{\cal D}}
\nc{\cE}{{\cal E}}
\nc{\cG}{{\cal G}}
\nc{\cH}{{\cal H}}
\nc{\cL}{{\cal L}}
\nc{\cO}{{\cal O}}
\nc{\cT}{{\cal T}}
\nc{\cN}{{\cal N}}
\nc{\rvac}[1]{|{\cal O}#1\rangle}
\nc{\lvac}[1]{\langle{\cal O}#1|}
\nc{\rvacb}[1]{|{\cal O}_\beta #1\rangle}
\nc{\lvacb}[1]{\langle{\cal O}_\beta #1 |}
\nc{\bb}{\bar{\beta}}
\nc{\bt}{\tilde{\beta}}
\nc{\ctH}{\tilde{\cal H}}
\nc{\chH}{\hat{\cal H}}
\nc{\1}{\aa}
\nc{\2}{\"{a}}
\nc{\3}{\"{o}}
\nc{\4}{\AA}
\nc{\5}{\"{A}}
\nc{\6}{\"{O}}
\nc{\al}{\alpha}
\nc{\g}{\gamma}
\nc{\Del}{\Delta}
\nc{\e}{\epsilon}
\nc{\eps}{\epsilon}
\nc{\lam}{\lambda}
\nc{\om}{\omega}
\nc{\Om}{\Omega}
\nc{\ve}{\varepsilon}
\nc{\mn}{{\mu\nu}}
\nc{\k}{\kappa}
\nc{\vp}{\varphi}

\nc{\advp}[3]{{\it  Adv.\ in\ Phys.\ }{{\bf #1} {(#2)} {#3}}}
\nc{\annp}[3]{{\it  Ann.\ Phys.\ (N.Y.)\ }{{\bf #1} {(#2)} {#3}}}
\nc{\apl}[3]{{\it  Appl. Phys. Lett. }{{\bf #1} {(#2)} {#3}}}
\nc{\apj}[3]{{\it  Ap.\ J.\ }{{\bf #1} {(#2)} {#3}}}
\nc{\apjl}[3]{{\it  Ap.\ J.\ Lett.\ }{{\bf #1} {(#2)} {#3}}}
\nc{\app}[3]{{\it Astropart.\ Phys.\ }{{\bf #1} {(#2)} {#3}}}  
\nc{\cmp}[3]{{\it  Comm.\ Math.\ Phys.\ }{{ \bf #1} {(#2)} {#3}}}
\nc{\cqg}[3]{{\it  Class.\ Quant.\ Grav.\ }{{\bf #1} {(#2)} {#3}}}
\nc{\epl}[3]{{\it  Europhys.\ Lett.\ }{{\bf #1} {(#2)} {#3}}}
\nc{\ijmp}[3]{{\it Int.\ J.\ Mod.\ Phys.\ }{{\bf #1} {(#2)} {#3}}}
\nc{\ijtp}[3]{{\it Int.\ J.\ Theor.\ Phys.\ }{{\bf #1} {(#2)} {#3}}}
\nc{\jmp}[3]{{\it  J.\ Math.\ Phys.\ }{{ \bf #1} {(#2)} {#3}}}
\nc{\jpa}[3]{{\it  J.\ Phys.\ A\ }{{\bf #1} {(#2)} {#3}}}
\nc{\jpc}[3]{{\it  J.\ Phys.\ C\ }{{\bf #1} {(#2)} {#3}}}
\nc{\jap}[3]{{\it J.\ Appl.\ Phys.\ }{{\bf #1} {(#2)} {#3}}}
\nc{\jpsj}[3]{{\it J.\ Phys.\ Soc.\ Japan\ }{{\bf #1} {(#2)} {#3}}}
\nc{\lmp}[3]{{\it Lett.\ Math.\ Phys.\ }{{\bf #1} {(#2)} {#3}}}
\nc{\mpl}[3]{{\it  Mod.\ Phys.\ Lett.\ }{{\bf #1} {(#2)} {#3}}}
\nc{\ncim}[3]{{\it  Nuov.\ Cim.\ }{{\bf #1} {(#2)} {#3}}}
\nc{\np}[3]{{\it  Nucl.\ Phys.\ }{{\bf #1} {(#2)} {#3}}}
\nc{\pr}[3]{{\it Phys.\ Rev.\ }{{\bf #1} {(#2)} {#3}}}
\nc{\pra}[3]{{\it  Phys.\ Rev.\ A\ }{{\bf #1} {(#2)} {#3}}}
\nc{\prb}[3]{{\it  Phys.\ Rev.\ B\ }{{{\bf #1} {(#2)} {#3}}}}
\nc{\prc}[3]{{\it  Phys.\ Rev.\ C\ }{{\bf #1} {(#2)} {#3}}}
\nc{\prd}[3]{{\it  Phys.\ Rev.\ D\ }{{\bf #1} {(#2)} {#3}}}
\nc{\prl}[3]{{\it Phys\ Rev.\ Lett.\ }{{\bf #1} {(#2)} {#3}}}
\nc{\pl}[3]{{\it  Phys.\ Lett.\ }{{\bf #1} {(#2)} {#3}}}
\nc{\prep}[3]{{\it Phys\. Rep.\ }{{\bf #1} {(#2)} {#3}}}
\nc{\prsl}[3]{{\it Proc.\ R.\ Soc.\ London\ }{{\bf #1} {(#2)} {#3}}}
\nc{\ptp}[3]{{\it  Prog.\ Theor.\ Phys.\ }{{\bf #1} {(#2)} {#3}}}
\nc{\ptps}[3]{{\it  Prog\ Theor.\ Phys.\ suppl.\ }{{\bf #1} {(#2)} {#3}}}
\nc{\physa}[3]{{\it  Physica\ A\ }{{\bf #1} {(#2)} {#3}}}
\nc{\physb}[3]{{\it  Physica\ B\ }{{\bf #1} {(#2)} {#3}}}
\nc{\phys}[3]{{\it Physica\ }{{\bf #1} {(#2)} {#3}}}
\nc{\rmp}[3]{{\it  Rev.\ Mod.\ Phys.\ }{{\bf #1} {(#2)} {#3}}}
\nc{\rpp}[3]{{\it Rep.\ Prog.\ Phys.\ }{{\bf #1} {(#2)} {#3}}}
\nc{\sjnp}[3]{{\it Sov.\ J.\ Nucl.\ Phys.\ }{{\bf #1} {(#2)} {#3}}}
\nc{\spjetp}[3]{{\it Sov.\ Phys.\ JETP\ }{{\bf #1} {(#2)} {#3}}}
\nc{\yf}[3]{{\it Yad.\ Fiz.\ }{{\bf #1} {(#2)} {#3}}}
\nc{\zetp}[3]{{\it Zh.\ Eksp.\ Teor.\ Fiz.\  }{{\bf #1}  {(#2)} {#3}}}
\nc{\zp}[3]{{\it Z.\ Phys.\ }{{\bf #1} {(#2)} {#3}}}
\nc{\ibid}[3]{{\sl ibid.\ }{{\bf #1} {#2} {#3}}}
\nc{\rf}[1]{(\ref{#1})}
\nc{\nn}{\nonumber \\*}
\nc{\bfB}{\bf{B}}
\nc{\bfv}{\bf{v}}
\nc{\bfx}{\bf{x}}
\nc{\bfy}{\bf{y}}
\nc{\vx}{\vec{x}}
\nc{\vy}{\vec{y}}
\nc{\oB}{\overline{B}}
\nc{\oI}{\overline{I}}
\nc{\oR}{\overline{R}}
\nc{\rar}{\rightarrow}
\nc{\ti}{\times}
\nc{\slsh}{\hskip-5pt/}
\nc{\sm}{Standard~Model~}
\nc{\w}{\omega}
\nc{\MP}{M_{\rm Pl}}
\nc{\tp}{t_{\rm Pl}}
\nc{\ave}{\bar{E}}

\renc{\min}{p_{\rm min}}
\renc{\max}{p_{\rm max}}
\nc{\pmin}{p_{\rm min}}
\nc{\pmax}{p_{\rm max}}
\nc{\fo}{f_0}
\nc{\foi}{f_{0,i}\,}
\nc{\fop}{f_0^P}
\nc{\fou}{f_0^U}
\def\sepand{\rule{14cm}{0pt}\and}
\nc{\eff}{{\rm eff}}
\nc{\MT}{M_{\rm T}}
\nc{\ML}{M_{\rm L}}
\nc{\kk}{\vek{k}}
\nc{\pp}{{\rm p}}
\nc{\cb}{critical bubble~}
\nc{\cbs}{critical bubbles~}
\nc{\scb}{subcritical bubble~}
\nc{\scbs}{subcritical bubbles~}
\begin{document}

{\title{{\hfill {{\small  TURKU-FL-P34-00
        }}\vskip 1truecm}
{\bf Q-ball collisions in the MSSM: gravity-mediated 
supersymmetry breaking}}

 
\author{
{\sc Tuomas Multam\" aki$^{1\dagger}$}\\
{\sl and}\\
{\sc Iiro Vilja$^{2}$ }\\ 
{\sl Department of Physics,
University of Turku} \\
{\sl FIN-20014 Turku, Finland} \\
\sepand
}
\maketitle}

\vspace{2cm}
\begin{abstract}
\noindent 
Collisions of non-topological solitons, Q-balls, are studied 
in a typical potential in the Minimal Supersymmetric Standard Model
where supersymmetry has been broken by a gravitationally coupled hidden sector.
Q-ball collisions are studied numerically on a two dimensional lattice 
for a range of Q-ball charges. Total cross-sections, as well as
cross-sections for fusion and charge-exchange are calculated. 
The average percentage increase in charge carried by the largest 
Q-ball after a collision is found to be weakly dependent 
on the initial charge.
\end{abstract}
\vfill
\footnoterule
{\small$^1$tuomas@maxwell.tfy.utu.fi,  $^2$vilja@newton.tfy.utu.fi\\}
{\small $\dagger$ Supported by the Finnish Graduate School in Nuclear and 
Particle Physics}
\thispagestyle{empty}
\newpage
\setcounter{page}{1}
\section{Introduction}
Stable non-topological solitons \cite{leepang}, Q-balls \cite{coleman},
can be present in several field theory models. In particular
the supersymmetric extensions of the Standard Model may
contain them.
A Q-ball is a coherent state of a complex
scalar field that carries a conserved charge,
typically a $U(1)$-charge. In the
sector of fixed charge a Q-ball is a ground state so that the
conservation of charge assures that the Q-ball is stable.
In the Minimal Supersymmetric Standard Model (MSSM) Q-balls
carrying lepton or baryon number are present
due to the existence of flat directions in the scalar sector
of the theory \cite{kusennko2,enqvist1}.

The cosmological significance of Q-balls can present itself
in many forms. Stable (or long living) Q-balls are natural candidates
for dark matter \cite{kusenko3} and the decay of Q-balls
can explain the baryon to dark matter ratio of the universe \cite{enqvistdm}.
Q-balls can also protect the baryon asymmetry from sphalerons 
at the electroweak phase transition and decaying Q-balls
may be responsible for the baryon asymmetry of the
universe \cite{enqvist1}. Furthermore, Q-balls can play an important role in 
considering the stability of neutron stars \cite{wreck}.

The mechanism by which supersymmetry (SUSY) is broken in the
theory is significant for the charges and stability
of Q-balls in the theory. If SUSY is broken by a gauge-mediated
mechanism, the baryon number carrying B-balls can have 
very large charges due to to the flatness of the potential.
Assuming that the charge is large enough, they can then be stable 
against decay into nucleons \cite{kusenko3}. If, however,
SUSY is broken by a gravitationally coupled hidden sector,
the potential is not completely flat but Q-balls 
may still exist due to radiative corrections \cite{enqvist1,
enqvistdm}. In this case the Q-balls can decay (evaporate \cite{cohen,
multamak}) into baryons or supersymmetric particles.

The formation of Q-balls from an Affleck-Dine (AD) condensate in the
early universe has been studied recently with numerical
simulations \cite{kawasaki1, kawasaki2}. In these simulations
both the gauge- and gravity-mediated SUSY breaking scenarios
have been considered. In both cases it was found that Q-balls
do form from the AD condensate. In the gravity-mediated case
it was especially noted that the formed Q-balls have non-zero
velocities and can hence collide with each other \cite{kawasaki2}.
Q-ball collisions were also simulated on a one dimensional lattice and 
it was found that Q-balls typically merge, exchange charge or pass 
through each other \cite{kawasaki2}.
Since the charge can change due to collisions, they may play
an important role in the determination of the Q-ball charge distribution 
after their formation. On the other hand the charge distribution is
important in evaluating the significance of Q-balls for the evolution
of the universe. It is hence worthwhile to study Q-ball collisions
in more detail. Q-ball collisions have also been studied previously in
various potentials in \cite{axenides}-\cite{battye} but not to our
knowledge in either the gauge- or gravity-mediated scenarios.

In this paper we have studied numerically collisions of Q-balls
in the gravity-mediated scenario on a two dimensional lattice. 
The gauge mediated scenario will be analyzed in a forthcoming paper.

\section{Q-ball solutions}
Consider a field theory with a U(1) symmetric scalar potential,
$U(\phi)$, with a global minimum at $\phi=0$. The complex scalar field
$\phi$ carries a unit quantum number with respect to the $U(1)$-symmetry.
The charge and energy of a given field configuration $\phi$ 
are given by \cite{leepang}
\be{charge}
Q={1\over i}\int (\phi^*\partial_t\phi-\phi\partial_t\phi^*)d^Dx
\ee
and
\be{energy}
E=\int [|\dot{\phi}|^2+|\nabla\phi|^2+U(\phi^*\phi)]d^Dx.
\ee
The single Q-ball solution corresponds to the minimum energy configuration 
at a fixed charge. If it is energetically more favourable to store
charge in a Q-ball compared to free particles, the Q-ball will be stable
against radiative decays into $\phi$-scalars.
Hence, for a stable Q-ball, condition
\be{stabcond}
E<mQ,
\ee
where $m$ is the mass of the $\phi$-scalar, must hold.

Minimizing the energy is straightforward and is easily done by using
Lagrange multipliers. The Q-ball solution can be shown to be of the form
\be{ansatz}
\phi(x,t)=e^{i\omega t}\phi(r),
\ee
where $\phi(x)$ is now time independent and real, $\omega$ is 
the Q-ball frequency, $|\omega|\in[0,m]$ and $\phi$ is spherically
symmetric.
The charge of a spherically symmetric Q-ball in
D-dimensions reads
\be{sphercharge}
Q=2\omega\int \phi(r)^2 d^Dr.
\ee
The equation of motion at a fixed $\omega$ is 
\be{eom}
{d^2\phi\over dr^2}+{D-1\over r}{d\phi\over dr}=\phi{\partial
U(\phi^2)\over\partial\phi^2}-\omega^2\phi.
\ee 
To obtain the Q-ball profiles we must solve (\ref{eom}) with
boundary conditions $\phi'(0)=0,\ \phi(\infty)=0$.

In the present paper we consider a potential of the form
\be{potential}
U(\phi) =m^2\phi^2(1-K \log({\phi^2\over M^2}))+\lambda\phi^{10},
\ee
where the parameter values are chosen to be $m=100$ GeV, $K=0.1$ and
$\lambda=\MP^{-6}$, where $\MP$ is the reduced Planck mass. 
This choice of potential corresponds to a D-flat direction in the 
full scalar potential in the MSSM where supersymmetry has been broken
by a gravitationally coupled hidden sector \cite{enqvist1}.
The large mass scale M is chosen such that the minimum is degenerate
\be{M}
M=({1\over 4}K m^2\lambda^{-1}\exp(-1-{4\over K}))^{1\over 8}\
\sim 10^{11}\GeV.
\ee

\begin{figure}[ht]
\leavevmode
\centering
\vspace*{55mm}
\includegraphics{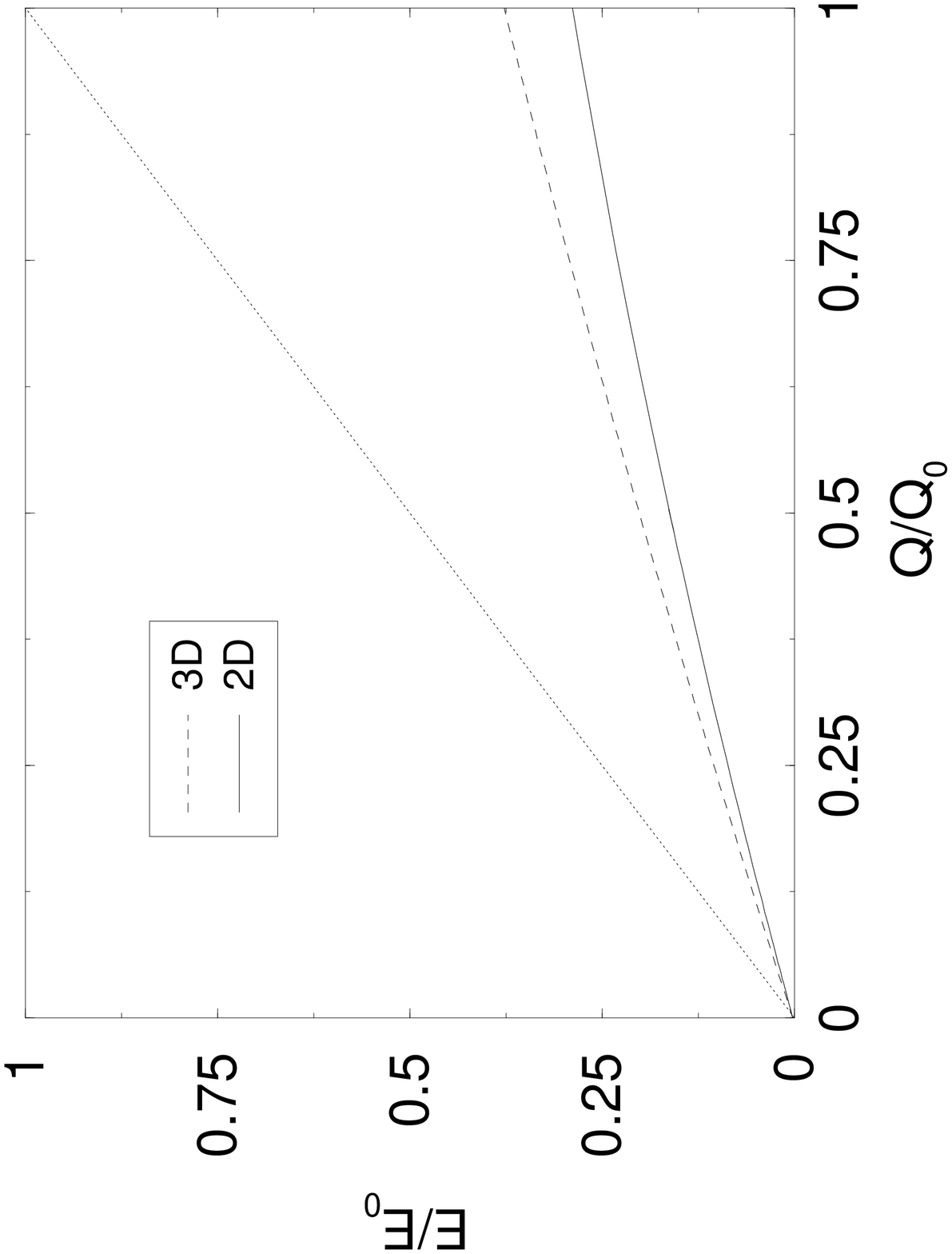}
\includegraphics{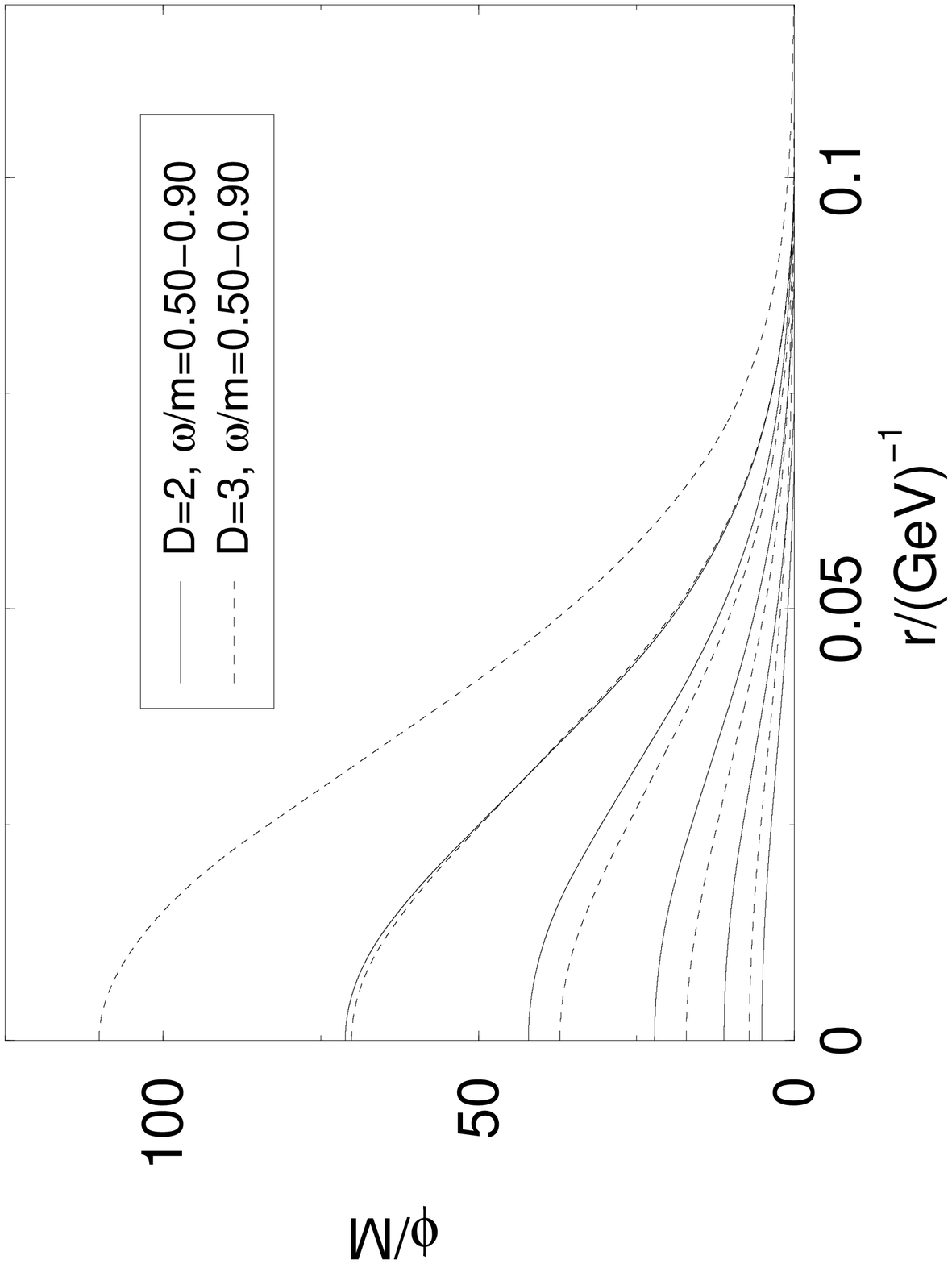}

\hspace{1cm}(a)\hspace{7cm}(b)
\caption{Q-ball energy as a function of charge and Q-ball profiles
in two and three dimensions.}
\label{evsq}
\end{figure}

We have calculated the charge and energy of Q-balls for different values
of $\w$. Energy vs. charge curves are shown in Figure \ref{evsq}(a).
The axis scales are chosen differently for two and three dimensions; 
for two dimensions, $Q_0=8000(M/\GeV)^2$, $E_0=8\times 10^{5}M^2\GeV^{-1}$
and for three dimensions, $Q_0=600(M/\GeV)^2$, 
$E_0=6\times 10^{4}M^2\GeV^{-1}$.
The dashed line is the stability line, $E=mQ$, that 
indicates that the Q-balls considered here are stable with
respect to scalar decays.
From the figure it can be seen that the energy vs. charge curves 
are of a similar shape in two and three dimensions. Q-ball profiles are 
plotted in Figure \ref{evsq}(b)
for different values of $\w$ in two and three dimensions. The
profiles appear very similar in these two cases. These figures
suggest that the collision processes calculated in two spatial
dimensions are likely to be similar to collisions in three spatial dimensions.
Here it is worth noting that, as from the equation of motion (\ref{eom})
can be seen, a non-zero dissipation term is present for dimensions
larger than one. This suggests that collisions in one dimension
may differ quite significantly from collision processes in higher
dimensions. This also seemed to be the case in the one dimensional
simulations we have done.

\section{Collisions}
We have studied collisions of Q-balls with equal charges in the 
potential (\ref{potential}) \cite{kuvat}.
The studied values of $\w$:s were $\w/m=0.50,\ 0.60,\ 0.75,\ 0.90,\ 0.99$. 
These values correspond to charges $1450,\ 645,\ 110,\ 12.6,\ 2.24$ in
units of $(M/\GeV)^2$ in the case with two spatial dimensions so that in terms 
of $\phi$-scalars the considered range of charges is $\sim 10^{24}-10^{26}$. 
For the same range of $\w$:s, the charges in the three dimensional
space are in the range $\sim 10^{23}-10^{26}$.

The relative phase of the Q-balls is also accounted for. 
By the relative phase, we mean the difference in individual 
phases at the point where the distance between Q-balls is at a 
minimum assuming there is no interaction between them. Position 
is defined as the point of
the maximum value of the amplitude of $\phi$.
The relative phase is allowed to have values in the range 
$0\leq\Delta\phi\leq2\pi$.
The impact parameter is also varied to study the cross-sections.
The Q-balls studied here are of the thick-wall type so that
there is no natural definition of the Q-ball size.
Therefore we have defined the size of a Q-ball by a Gaussian fit; we fit
a Gaussian $\phi=A e^{-Br^2}$ to the profiles and define
the radius of the ball as $R=B^{-{1\over2}}$.
The cross-sections quoted here are three dimensional cross-sections
with the interaction radius taken from the two-dimensional simulations.

Collisions were simulated on a 2+1 -dimensional lattice. The
lattice size typically used was $\sim200^2$ with continuous boundary 
conditions. A 9-point Laplacian operator and a step size of $5\times 
10^{-3}$ was used in all calculations. 
Collisions were studied for different initial velocities of Q-balls,
$v=10^{-3}$ and $v=10^{-2}$.

\subsection{Numerical Results}
\begin{figure}[ht]
\leavevmode
\centering
\vspace*{110mm}
\includegraphics{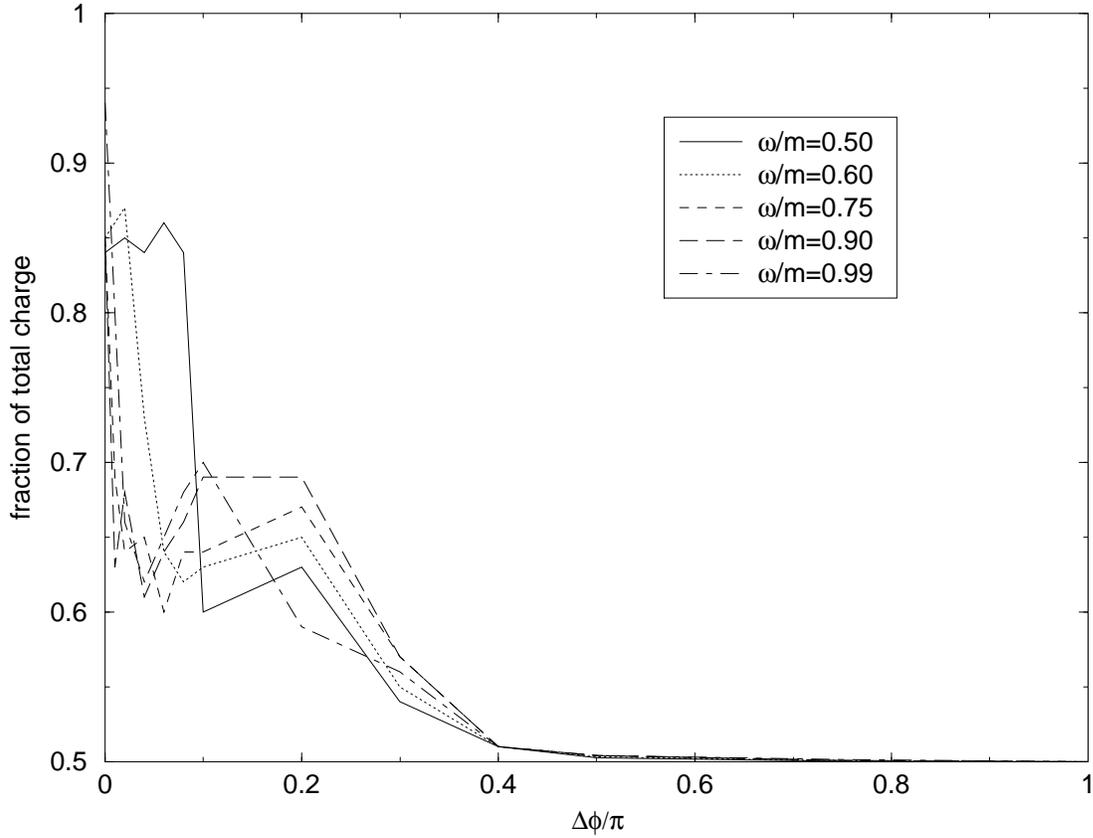}
\caption{The fraction of initial charge in the largest Q-ball after a
collision for different values of $\Delta\phi$, $v=10^{-3}$.}
\label{qfracs}
\end{figure}

\begin{figure}[ht]
\leavevmode
\centering
\vspace*{110mm}
\includegraphics{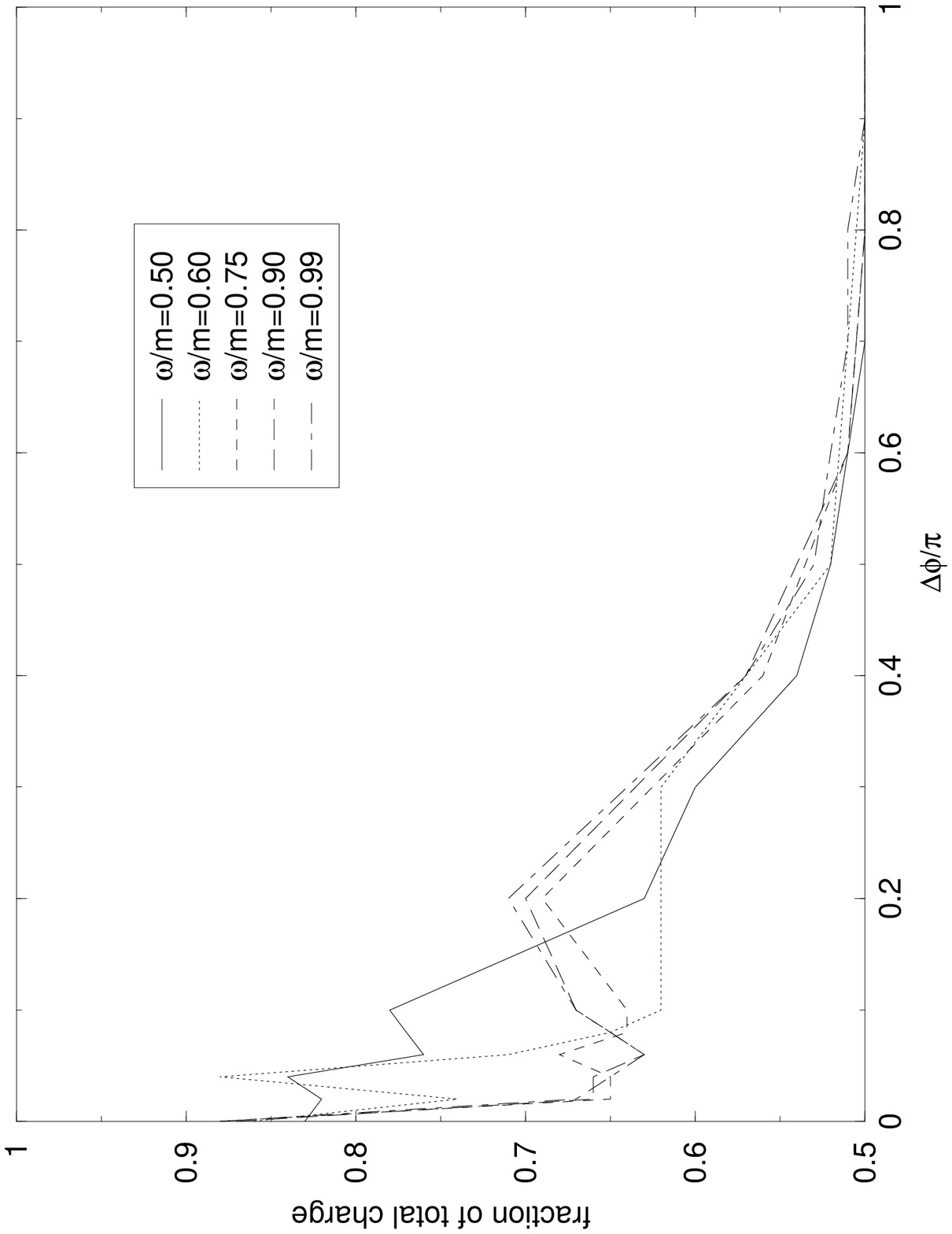}
\caption{The fraction of initial charge in the largest Q-ball after a
collision for different values of $\Delta\phi$, $v=10^{-2}$.}
\label{qfracs2}
\end{figure}

Collisions can roughly be divided into three types; fusion, charge
exchange and elastic scattering. Fusion is defined as a process where
most of the initial charge is in a single Q-ball after
the collision and the rest of the charge is lost
either as radiation or as small Q-balls.
By charge exchange we mean a process where Q-balls exchange
some of their charge while the total amount of charge carried by the
two balls is essentially conserved. An elastic scattering is defined
to be a process where less than $1\%$ of the total charge is exchanged.
After the collision the ratio
of the charge in the largest Q-ball to the total initial charge 
as a function of the relative phase has been plotted in Figures
\ref{qfracs} and \ref{qfracs2}.
From the figures the two different types of processes can be
distinguished. In a fusion process typically $\sim 10-20\%$ of the 
initial charge is lost as radiation and small Q-balls and the rest 
of the charge is in a single Q-ball. On the other hand, if 
charge is exchanged the larger
Q-ball carries usually less than $70\%$ of the total charge.
As from Figs. \ref{qfracs} and \ref{qfracs2} can be seen,
fusion occurs generally only when the relative phase is small
and is more likely to occur with smaller $\w$. 
The amount of exchanged charge decreases substantially with increasing 
relative phase. 
Increased velocity does not seem to have a large effect for
the range of $\w$ where fusion occurs. However, more charge is exchanged
between balls of equal size when the initial velocity is larger.
The relative changes in size are weakly dependent on $\w$ 
(in both cases the standard deviation is less than $1\%$) 
and hence on the size of the Q-balls.
Averaging over the relative phase (assuming a random distribution
for the $\Delta\phi$:s) and $\w$:s, the relative change
in the size of a Q-ball is $10\%$ for $v=10^{-3}$ and
$14\%$ for $v=10^{-2}$. 

\begin{figure}[ht]
\leavevmode
\centering
\vspace*{110mm}
\includegraphics{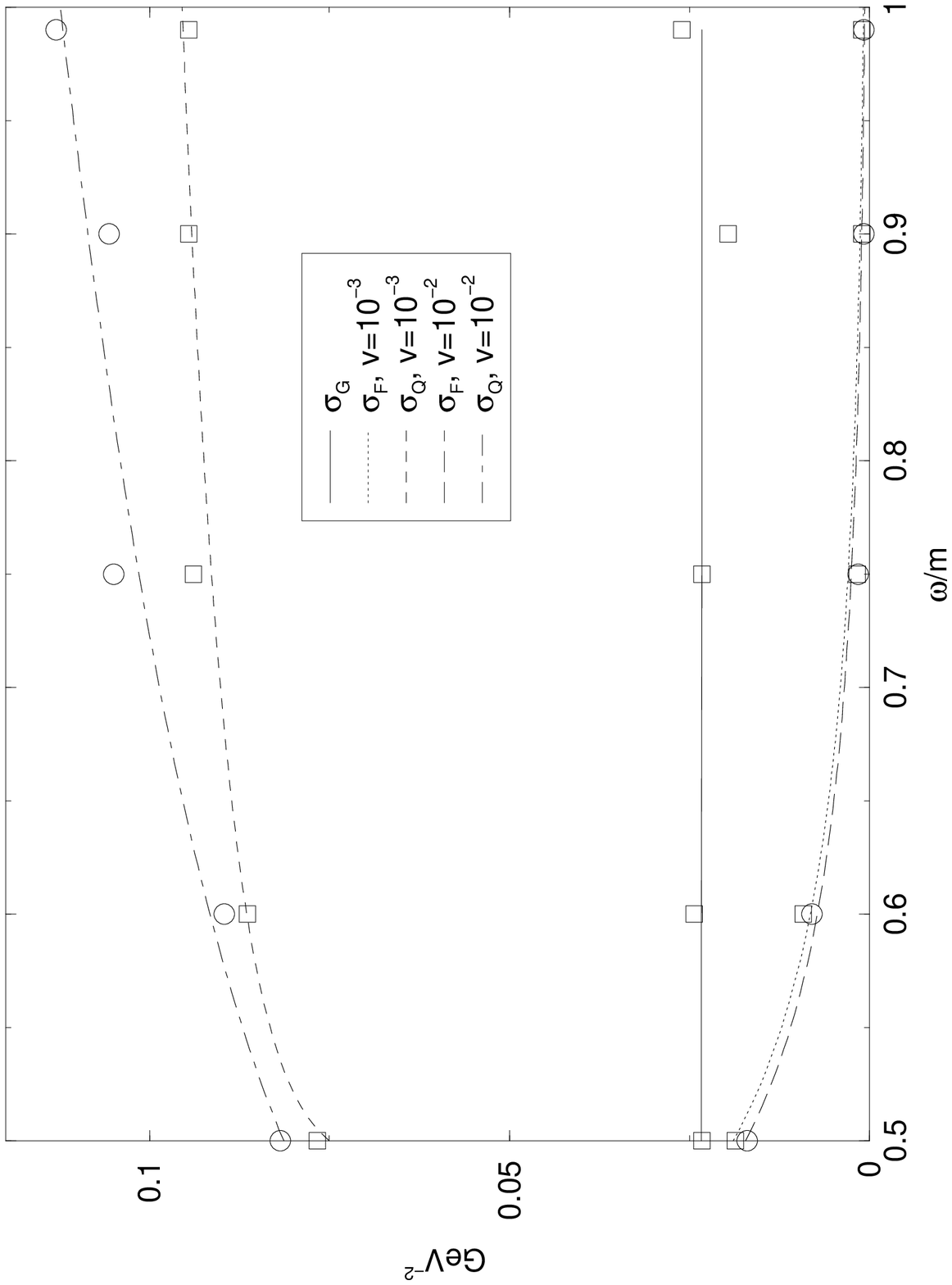}
\caption{Radius of the Q-ball and average fusion and charge exchange 
cross-sections for different values of $\w$.}
\label{fusion}
\end{figure}

We are now ready to calculate the fusion cross-section, $\sigma_{\rm{F}}$,
and the cross-section for the charge exchange, $\sigma_{\rm{Q}}$,
as a function of $\w$.
These are plotted in Figure \ref{fusion} with the geometrical
cross-section, $\sigma_{\rm{G}}$.
Clearly the fusion cross-section is strongly dependent on $\w$;
larger Q-balls fuse more easily than smaller ones. This effect
is clearly not explained by the different geometrical sizes of the
Q-balls as can be noted from the geometrical cross-section.
The fusion cross-section decreases with increasing $\w$
because the Q-balls with higher $\w$ have more regions with differing 
relative phases. The field dynamics cannot then even out the relative
phase differences quickly enough to keep the colliding balls together. 
From the simulations it can be seen that balls with larger $\w$ are
less likely to fuse than balls with the same phase difference but
smaller $\w$.
The effect of increasing the initial velocity on the cross-sections
can also be noted from Fig. \ref{fusion}. The fusion cross-section
is slightly decreased as velocity increases while the charge exchange
cross-section increases. The increase in $\sigma_Q$ is due to the fact 
that now the Q-balls have more kinetic energy to overcome the
repulsion resulting from the relative phase difference.

The total cross-section including all the cases
\ie when the balls fuse, exchange charge or scatter elastically, 
is also dependent on $\w$, but only weakly. Averaging over the
$\w$:s, $\sigma_{\textrm{tot}}=0.27\pm 0.01\GeV^{-2}$ ($v=10^{-3}$)
and $\sigma_{\textrm{tot}}=0.19\pm 0.01\GeV^{-2}$ ($v=10^{-2}$).

We have also studied Q-ball collisions with larger initial velocities
for a more limited set of parameter values.
When the initial velocity is increased to $v=10^{-1}$, the fusion
cross-section is reduced significantly from its value when $v=10^{-3}$.
Furthermore, at such high velocities we also see processes where
the Q-balls pass through each other essentially without exchanging 
any charge. This is a similar process that was reported to occur
in one dimension in \cite{kawasaki2} and which we have also 
observed in our one dimensional simulations.

Charge exchange also affects the final velocities of the Q-balls. 
As charge is exchanged the speed of the Q-balls typically increase.
The final velocity of the smaller ball can be quite large; we 
have often noted final velocities ten times larger than the 
initial velocity.

\section{Conclusions}

In this paper we have studied Q-ball collisions in the
MSSM with supersymmetry broken by a gravitational hidden
sector. For the studied range of charges the total cross-section
was found to be approximately constant.
The cross-section for fusion, $\sigma_{\rm{F}}$, appeared to be smaller
than the geometrical cross-section, $\sigma_{\rm{G}}$, whereas
the cross-section for charge exchange, $\sigma_{\rm{Q}}$, was
larger than $\sigma_{\rm{G}}$. In a collision it is hence more probable
that a charge exchanging process occurs rather than a fusion process.
This probability increases with increasing $\w$ (or with decreasing
charge).
Averaging over the fusion and charge exchanging processes
the average charge increase of the largest Q-ball emerging from
a collision was found to be approximately constant. For the considered 
range of charges and velocities it was $\sim 10\%\ (v=10^{-3})$ and
$\sim 14\%\ (v=10^{-2})$.

In a cosmological context, Q-ball collisions may have a significant 
effect on the charge distribution of Q-balls. 
Clearly for collisions to be important the number density of Q-balls
must be high enough and the balls must have large enough velocities for
the rate of interaction to be significant. 
In the early universe this obviously means that
the interaction rate must be larger than the Hubble rate. If collisions
typically do occur the resulting charge distribution can then be altered
by the fusion and charge exchange processes. Based on the results
presented in this paper, the relative phase, size and the initial
velocity of the balls then play important roles in studying the 
evolution of the Q-ball charge distribution. 

If the balls that are formed from the AD condensate are in the same phase,
fusion processes will dominate and the average size of a Q-ball
grows substantially in a collision. Since most of the charge is left in
the remaining ball and the rest is in the form of several small,
quickly evaporating \cite{multamak}, Q-balls
and radiation, the number density of Q-balls reduces rapidly. Collisions
can therefore freeze the Q-ball distribution quickly in the early phases
of the universe. If, on the other hand, the
phases are randomly distributed the probability for fusion is greatly
reduced and the distribution will not change as significantly as
in the previous case. Collision processes can then also continue for a
longer period of time.

The typical size of Q-balls is obviously an important factor. The
total scattering cross-section depends quite weakly on the size of the
Q-balls but the fusion and charge exchange cross-sections do have a strong
$\w$-dependence. 

The initial velocity of the Q-balls is also significant in the
evolution of the Q-ball distribution. A larger velocity means that 
the interaction rate is increased but on the other hand if the initial
velocity is too large the cross-sections decrease due to a decreased
interaction time. Collisions can also
significantly change the velocities of the Q-balls so that an initially
uniform velocity distribution can be spread out by the collision
processes.

The effect of collisions can be important in deciding the exact role
and significance of Q-balls in the evolution of the universe.
In determining their importance on cosmology, more information is needed
about the Q-ball distribution after their formation and also about 
the effects of collisions on the initial distribution.
To quantify the effects of the different collision processes described
in this paper, a more detailed analysis is needed which gives motivation 
for future work.

\vspace{1cm}
\noindent{\bf Acknowledgements.}
We thank K. Enqvist for discussions and the Center for Scientific
Computing for computation resources. This work has been supported 
by the Academy of Finland.


\end{document}